# Structure retrieval in liquid-phase electron scattering

Jie Yang[1,2]*, J. Pedro F. Nunes[3], Kathryn Ledbetter[2,4], Elisa Biasin[1,2], Martin Centurion[3], Zhijiang Chen[1], Amy A. Cordones[2], Christ Crissman[1], Daniel P. Deponte[1], Siegfried H. Glenzer[1], Ming-Fu Lin[1], Mianzhen Mo[1], Conor D. Rankine[5], Xiaozhe Shen[1], Thomas J. A. Wolf[2], Xijie Wang[1]*

[1]*SLAC National Accelerator Laboratory, Menlo Park, California, 94025, USA.*

[2]*Stanford PULSE Institute, SLAC National Accelerator Laboratory, Menlo Park, California, 94025, USA.*

[3]*Department of Physics and Astronomy, University of Nebraska–Lincoln, Lincoln, Nebraska, 68588, USA.*

[4]*Physics Department, Stanford University, Stanford, California, 94305, USA.*

[5]*School of Natural and Environmental Sciences, Newcastle University, Newcastle upon Tyne, NE1 7RU, UK.*

The molecular structure of liquids has been widely studied using neutron and X-ray scattering. Recent ultrathin liquid sheet technologies have enabled electron scattering on liquid samples. The data treatment and of liquid-phase electron scattering has been mostly reliant on methodologies developed for gas electron diffraction, in which theoretical inputs and empirical fittings are often needed to account for the atomic form factor and remove the inelastic scattering background. The accuracy and impact of these theoretical and empirical inputs has not been benchmarked for liquid-phase electron scattering data. In this work, we present a mathematically rigorous data treatment method that requires neither theoretical inputs nor empirical fittings. The merits of this new method are illustrated through the retrieval of real-space molecular structure from experimental electron scattering patterns of liquid water, carbon tetrachloride, chloroform, and dichloromethane.

## I. INTRODUCTION

Scattering provides a direct view of molecular structure in matter. For samples in the liquid phase, X-ray and neutron scattering have been the two methods of choice [1-4], despite requiring large-scale facilities. Electron scattering, on the other hand, can be carried out using miniaturized and inexpensive table-top instruments in university labs [5-8]. However, the shallow penetration depth of electrons (<1 μm) compared to hard X-rays (>100 μm) and neutrons (>1 cm) have largely hindered its applicability in liquid phase samples. To avoid the loss of information due to multiple scattering, sample thicknesses on the order of a few hundred nm are required for high energy (~100 keV and above) electrons [9]. Since the 1970s, liquid electron scattering (LES) experiments using evaporating films [10], vapor deposition [11, 12], nanofluidic cells [13, 14], and free-flowing ultrathin liquid sheet jet [15, 16] have been developed with different levels of success. Most existing works have focused on the development of instrumentation, whereas the data treatment methods have been largely adopted

*Corresponding author: jieyang@slac.stanford.edu; wangxj@slac.stanford.edu

J. Y. and J. P. F. N. contributed equally to this work.

from gas electron diffraction (GED). These methods often rely on theoretical inputs and empirical fittings that are inherited from GED and not benchmarked for liquid samples.

In this work, we introduce an alternative data analysis method, charge-pair-distribution-function (*CPDF*), which is able to deliver real-space structural information directly from scattering patterns without relying on any theoretical inputs or empirical fittings. We present high-quality experimental LES data for water, $CCl_4$, $CHCl_3$, and $CH_2Cl_2$. These data were recorded at the SLAC MeV-UED facility using gas-accelerated and converging liquid jets capable of producing free-flowing ultrathin liquid sheets [15-17]. A schematic drawing of the experimental setup is shown in Fig.1(a) and a sample scattering pattern of $CCl_4$ is shown in Fig.1(b).

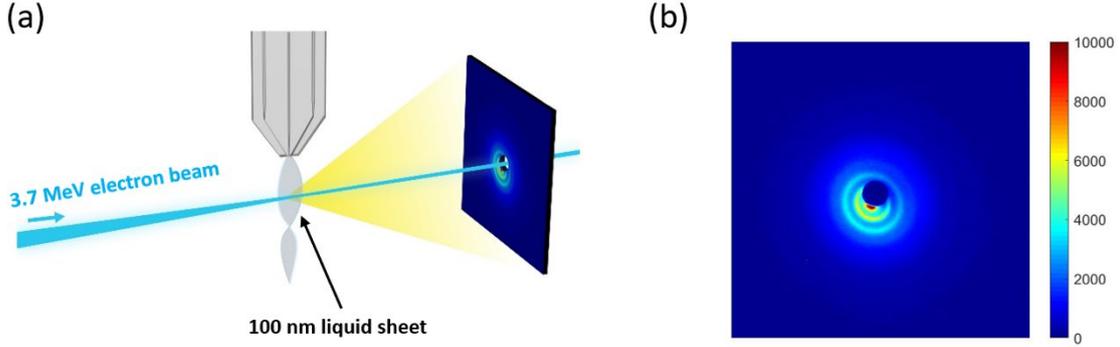

FIG. 1. Experiment illustration. (a). Schematic illustration of the experimental setup. (b). Scattering pattern of liquid $CCl_4$. The pattern is deliberately off-center in order to access the small-angle scattering. The color bar represents the detector counts for a 1-second exposure.

## II. Data Treatment Theory

In this section, we will first review the theory of electron scattering and the conventional pair distribution function (*PDF*) data treatment method commonplace in GED and then present *CPDF* method developed for LES.

Under the 1$^{st}$ Born approximation, the total electron scattering intensity (including both elastic and inelastic components) for a molecule can be written as [18]:

$$I(\vec{Q}) = \frac{1}{Q^4} \int \sum_u \sum_v Z_u Z_v P_{uv}(\vec{r}) \exp(i\vec{Q}\cdot\vec{r})d\vec{r} \quad (1)$$

Here $\vec{Q}$ is the momentum transfer, indices *u* and *v* run over all charged particles (nuclei and electrons), and $Z_u$ is the charge of the $u^{th}$ particle in units of elementary charge (−1 for electrons). $P_{uv}(\vec{r})$ represents the probability distribution of finding *uv* pairs at position $\vec{r}$. When $u = v$, $P_{uv}(\vec{r}) = \delta(\vec{r})$, the Dirac delta function. To keep the format concise, a constant $4\gamma^2/a_0^2$ is ignored in Eq. (1), where $\gamma$ is the Lorentz factor and $a_0$ is the Bohr radius. Eq. (1) was initially derived for gas phase molecules, where the space between molecules are much larger than the coherence length of the probe electrons and thus can be considered as isolated molecules in GED experiments. For liquid samples, the double sum in Eq. (1) is no longer limited to within a single molecule, but need to be performed over all charged particles within the coherent volume of the probe electron beam.

The total scattering intensity, $I(\vec{Q})$, can be separated into elastic and inelastic component [18-20],

$$I(\vec{Q}) = I_{elastic}(\vec{Q}) + I_{inelastic}(\vec{Q}) \quad (2)$$

where,

$$I_{elastic}(\vec{Q}) = Q^{-4} |\sum_u Z_u e^{i\vec{Q}\cdot\vec{R}_u} - F_1(\vec{Q})|^2 \quad (3)$$

$$I_{inelastic}(\vec{Q}) = Q^{-4}[n + F_2(\vec{Q}) - |F_1(\vec{Q})|^2] \quad (4)$$

in which $\vec{Q}$ is the momentum transfer, $\vec{R}_u$ is the position of the $u^{th}$ nucleus, $n$ is the number of electrons in the target, $F_1$ and $F_2$ are the Fourier transform of the one-electron and two-electron density, defined as

$$F_1(\vec{Q}) = \int e^{i\vec{Q}\cdot\vec{r}} \rho(\vec{r}) d\vec{r} \qquad (5)$$

$$F_2(\vec{Q}) = \int e^{i\vec{Q}\cdot(\vec{r}-\vec{r}\,')} \rho^{(2)}(\vec{r},\vec{r}\,') d\vec{r} d\vec{r}\,' \qquad (6)$$

where $\rho(\vec{r})$ and $\rho^{(2)}(\vec{r},\vec{r}\,')$ are the one-electron and two-electron reduced density operator, respectively.

Directly applying Eq. (3-6) for scattering pattern prediction is computationally very expensive. A convenient approximation that is often used in GED data processing is the independent atom model (IAM), in which electrons are assumed to be distributed around their parent nuclei as if these were isolated atoms and thus ignoring the electron redistribution due to the formation of chemical bonds. Under this approximation, the elastic component can be separated into atomic and molecular parts [21, 22],

$$I_{elastic}(\vec{Q}) = I_{at}(Q) + I_{mol}(\vec{Q}) \qquad (7)$$

where

$$I_{at}(Q) = \sum_u |f_u(Q)|^2 \qquad (8)$$

$$I_{mol}(\vec{Q}) = \sum_u \sum_{v \neq u} f_u^*(Q) f_v(Q) \exp(i\vec{Q}\cdot\vec{r}_{uv}) \qquad (9)$$

where $f_u$ is the atomic form factor (AFF) of elastic electron scattering for the $u^{th}$ atom and $\vec{r}_{uv}$ is a vector between the $u^{th}$ and $v^{th}$ nuclei. The relation between AFF for elastic electron scattering and AFF for elastic X-ray scattering is given by,

$$f_u(Q) = Q^{-2} |Z_u - f_u^{X-ray}(Q)| \qquad (10)$$

For liquid or gas samples, as molecules are randomly oriented, $I_{mol}$ becomes isotropic and Eq. (9) can be reduced to:

$$I_{mol}(Q) = \sum_u \sum_{v \neq u} f_u^*(Q) f_v(Q) \frac{\sin(Qr_{uv})}{Qr_{uv}} \qquad (11)$$

Under IAM, the inelastic component only depends on the type of atom, which can be written as

$$I_{inelastic}(Q) = Q^{-4} \sum_u S_u(Q) \qquad (12)$$

where $S_u$ is the inelastic form factor (IFF) for the $u^{th}$ atom. Note $S_u$ is identical to the IFF for X-ray scattering, so that the inelastic component for electron and X-ray scattering are only differ by the $Q^{-4}$ factor.

Among all the components, typically only $I_{mol}$ contains structural information of the target system. The rapid drop in $I_{mol}$ amplitude, imparted a $Q^{-5}$ dependency is typically overcome through the use of modified scattering intensity curves, $sM(Q)$, in which each atom pair distance appears sinusoidal without damping across the Q range.

$$sM(Q) = \frac{I_{mol}(Q)}{I_{at}(Q)} Q \qquad (13)$$

The PDF can then be calculated from $sM(Q)$ using a sine transform

$$PDF(r) = \int sM(Q) \sin(Qr) dQ \qquad (14)$$

In principle, PDF analysis must always be preceded by the removal of inelastic and atomic scattering contribution from the experimental data, a process which often requires theoretical inputs and empirical fittings, even in the well-established GED data analysis [22]. In X-ray scattering, for example, it has been found that even high-level quantum chemical calculations are insufficient to adequately remove the inelastic scattering background of liquid water [23]. Moreover, after getting $I_{mol}(Q)$, theoretical AFFs are still needed (Eq. (8) and (13)) in order to calculate the $sM(Q)$. For liquid, AFFs under IAM often need to be modified, for example, to incorporate the dipole moment in the case of a polar molecule such as water [4].

Here we present an alternative method to retrieve real-space information directly from LES patterns with neither background removal nor AFFs. Directly from Eq.(1) above, the scattering intensity from a randomly oriented molecular ensemble can be represented by

$$I(Q) = \frac{1}{Q^4} \int \sum_u \sum_v Z_u Z_v P_{uv}(r) \frac{\sin(Qr)}{Qr} dr \qquad (15)$$

We define a charge-pair distribution function $CPDF(r)$, as:

$$CPDF(r) = \sum_u \sum_v Z_u Z_v P_{uv}(r) \qquad (16)$$

It includes all nucleus-nucleus, nucleus-electron, and electron-electron pairs, with their amplitude and sign being determined by the product of the charge of the two particles.

Substituting Eq. (16) into Eq. (15), we obtain:

$$Q^5 I(Q) = \int_0^\infty CPDF(r) \frac{\sin(Qr)}{r} dr \qquad (17)$$

the sine transform of which yields

$$CPDF(r) = r \int_0^\infty Q^5 I(Q) \sin(Qr) dQ \qquad (18)$$

Since only a finite range of $Q$ is measured in any experiment, $I$ is multiplied by a damping term, $e^{-\alpha Q^2}$, to avoid edge effects during the transform.

$$CPDF(r) = r \int_0^{Q_{MAX}} Q^5 I(Q) e^{-\alpha Q^2} \sin(Qr) dQ \qquad (19)$$

This edge effect mitigation has been widely used in X-ray and electron scattering experiments, see for example Ref.[24]. The damping term is equivalent to a Gaussian smoothing for $CPDF(r)$ in the real space.

Note that the formula of the $CPDF(r)$ in Eq. (19) is a mathematically rigorous derivation from Eq. (1) without invoking approximations such as the IAM. In addition, it requires neither empirical fitting nor theoretical inputs, because no background removal or dividing by AFFs is performed.

Dividing by AFFs during transformation (Eq.13) is equivalent to a real-space deconvolution that replaces the "form" of atom with a point-like nucleus. For X-ray scattering, the "form" of an atom is a fuzzy electron cloud. For electron scattering, the "form" of an atom is a point-like positive charge (nucleus) embedded in a fuzzy negative cloud (electrons). For this reason, despite not using the AFFs, the $CPDF$ is still able to return sharp nucleus-nucleus pairs. However, this introduces negative shoulders that accompany the positive nucleus-nucleus pair peak, which arise from the fuzzy nature of electron clouds. Although this could be seen as an inconvenience in understanding liquid structure, it also could enhance the contrast of the nuclei, as per demonstrated in Fig. 6 in this work. In addition, the embedded electron-nucleus and electron-electron pairs may lead to a slight distortion of the nucleus-nucleus peaks. In this case, the peak positions will no longer directly reflect the position of the internuclear distances. We estimate such a shift to be within 0.05 Å. A detailed analysis of contributions from nucleus-nucleus, nucleus-electron and electron-electron pairs is given in Section IV A.

## III. EXPERIMENTAL DATA

### A. *CPDF* of liquid water

The experimental and simulated scattering patterns of liquid water, along with the key steps in calculating the *CPDF* ($I$, $Q^5 I$, $Q^5 I e^{-\alpha Q^2}$ and $CPDF(r)$) are shown in Fig. 2(a)-(d). The blue curve was measured from a gas-accelerated liquid jet with a thickness of ~100 nm and an electron transmission of 88%. The black curve was measured from a converging liquid jet with a thickness of ~650 nm and an electron transmission of only 40%. The jet thickness was measured, in each case, using an optical interferometer [16]. A classical molecular dynamics (MD) simulation for liquid water was performed using the GROMACS software suite with the TIP4P-Ew force-field[25]. The scattering simulation was performed under the IAM, as outlined in the appendix. The discrepancy between simulated and experimental $Q^5 I$ signal (Fig. 2(b)-(c)), with the experimental signal being significantly higher than the simulated signal at high $Q$. This discrepancy is likely due to the multiple-scattering background or residue camera background being amplified by the $Q^5$ factor. Nevertheless, this smooth background only contributes at small distances, and very good agreement was found between the experimental and simulated *CPDF* for $r > 2$ Å (Fig. 2(d)). An additional advantage of this method is that it is insensitive to low $Q$ scattering. In electron scattering, it is known that the IAM tends to overestimate low $Q$ scattering, as it fails to account for binding and correlation effects [26]. In addition, it is challenging to acquire experimentally electron scattering patterns down to $Q = 0$. In this work the minimum $Q$ value measured is between 0.3 and 0.5 Å$^{-1}$. Both the IAM inaccuracy and the missing data in low $Q$ have negligible contribution to the *CPDF*, since the $Q^5$ factor significantly reduces the weight of low-$Q$ scattering in the sine transform.

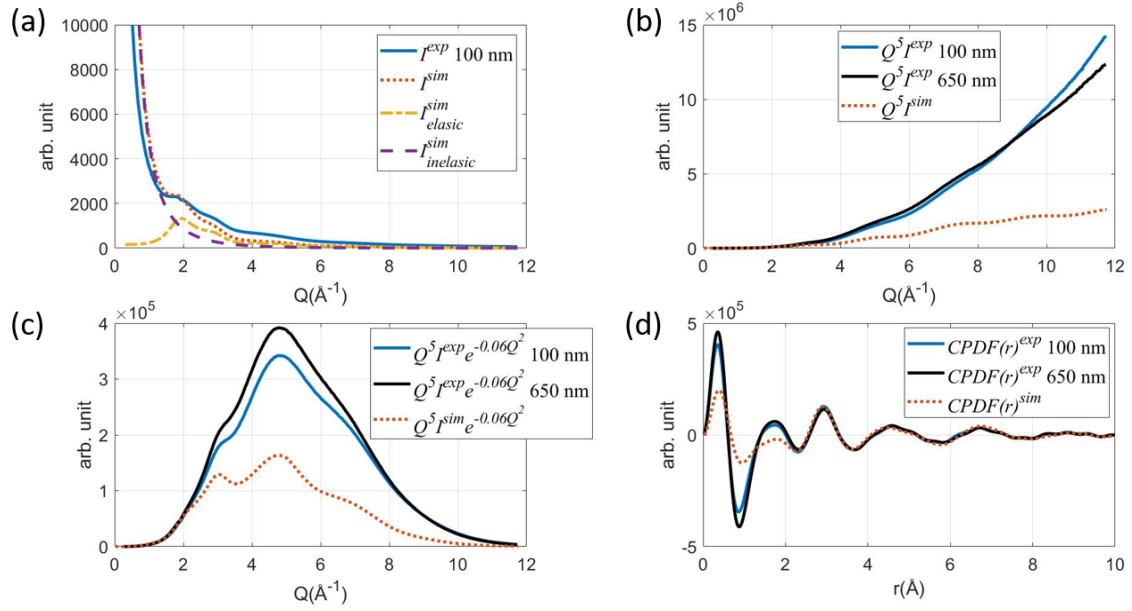

FIG. 2. Generating *CPDF* from liquid electron scattering patterns. Experimental 100 nm sheet jet (blue solid), 650 nm sheet jet (black solid) and simulated (red dotted) (a) $I$, (b) $Q^5I$, (c) $Q^5Ie^{-\alpha Q^2}$, (d) *CPDF(r)* for water. Elastic (yellow dash-dotted) and inelastic (purple dashed) components from simulation are also shown in part (a). Simulations were performed under the IAM, and the *CPDF* was calculated using Eq. (19) with a damping factor $\alpha = 0.06$.

The large difference in electron transmission between the two datasets suggests very different jet conditions and background levels. Nevertheless, the *CPDF* of the two datasets are still very similar, with the majority of differences concentrated at the small distances. This comparison shows that our method is relatively insensitive to the thickness of the jet – in the range between 100 and 650 nm, at least.

To further understand the simulated *CPDF* of water, we separate it into three components—elastic intraatomic contributions $CPDF_{intra}$, elastic interatomic contributions $CPDF_{inter}$, and inelastic contributions $CPDF_{ine}$. The IAM simulation of three components are shown in Fig. 3(a). $CPDF_{intra}$ covers the interference of charge pairs within each atom, $CPDF_{inter}$ covers the interference of charge pairs across different atoms, and $CPDF_{ine}$ covers the correction from electron correlation effect. The definition of each component is given in Section B in the appendix. For $r > \sim 2$ Å the $CPDF_{inter}$ contribution dominates and $CPDF_{intra}$ and $CPDF_{ine}$ both contribute negligibly. This feature is originated from the underlying physics that both intraatomic and inelastic component are short-ranged, and explains how the *CPDF* shown in Fig. 2(d) captures all nucleus pairs for $r > \sim 2$ Å without removing any background. A more detailed analysis of each component in *CPDF* is given in section IV A. Nevertheless, background removal is still required in order to retrieve nuclear pairs with $r < 2$ Å. In water, the two shortest nuclear pair distances, the bonded O-H (~1.0 Å) and the hydrogen-bonded O···H (~1.9 Å), are obscured by $CPDF_{intra}$ and $CPDF_{ine}$. Figure 3(b) shows the scattering pattern for each term after being multiplied by $Q^5$, with the molecular component, $Q^5I_{inter}$, oscillating around zero while $Q^5I_{intra}$ and $Q^5I_{ine}$ increase monotonically and smoothly. Therefore, a smooth background fitting to the $Q^5I$ curve could potentially remove $Q^5I_{intra}$ and $Q^5I_{ine}$, isolating the $Q^5I_{inter}$ contribution. Here we use a simple method to remove this smooth background: a low-order polynomial (3rd order here) is fitted to the full range (0.3 < $Q$ < 11.8 Å$^{-1}$) of experimental $Q^5I$, and $Q^2$ is multiplied to the range 0 < $Q$ < 1 Å$^{-1}$ to make the background smoothly go to zero at $Q = 0$. The fitted background is shown in Fig. 3(c), and the background-removed experimental data is plotted together with the simulated

$Q^5I_{inter}$ in Fig. 3(d). The polynomial fitting is a simple yet rudimentary background removal method that does not require any prior knowledge about the target system. The *CPDF*s, as retrieved from experimental data with background removal, are shown in Fig. 3(e) together with simulation. The peaks in *CPDF* appear sharper in Fig. 3(e) than Fig. 2(d), this is because that after background removal, the $Q^5I$ is no longer increasing rapidly with $Q$, and the damping factor $\alpha$ can be much relaxed. We used $\alpha=0.03$ for Fig. 3(e) and $\alpha=0.06$ for Fig. 2(d). While the background removal helps to reveal the two peaks under 2 Å, it also introduces certain artifacts, including, noticeably, a negative shift for the region 2 Å < $r$ < 4 Å and a positive shift for the peak at 4.5 Å. It also appears to negatively shift the peak at around 2 Å, leading to an underestimation of the length of the hydrogen bond in water. Therefore, for all the $r$ > 2 Å peaks, we give higher credence to the *CPDF* without background removal, and only use background removal to retrieve peaks under 2 Å. More advanced background removal methods, such as a $\chi^2$ background fitting based on zero-crossings of $I_{inter}$ [6], smooth background method [27], and earlier empirical methods [28], among others, have been utilized in gas-phase electron diffraction (GED) data analysis. Nevertheless, these methods typically require a certain amount of *a priori* knowledge or assumptions to be made about the system under study. The applicability of these methods to LES are subject to future studies. In the rest of this work, we will focus on the *CPDF* without background removal, and only focus on the $r$ > 2 Å pair distances.

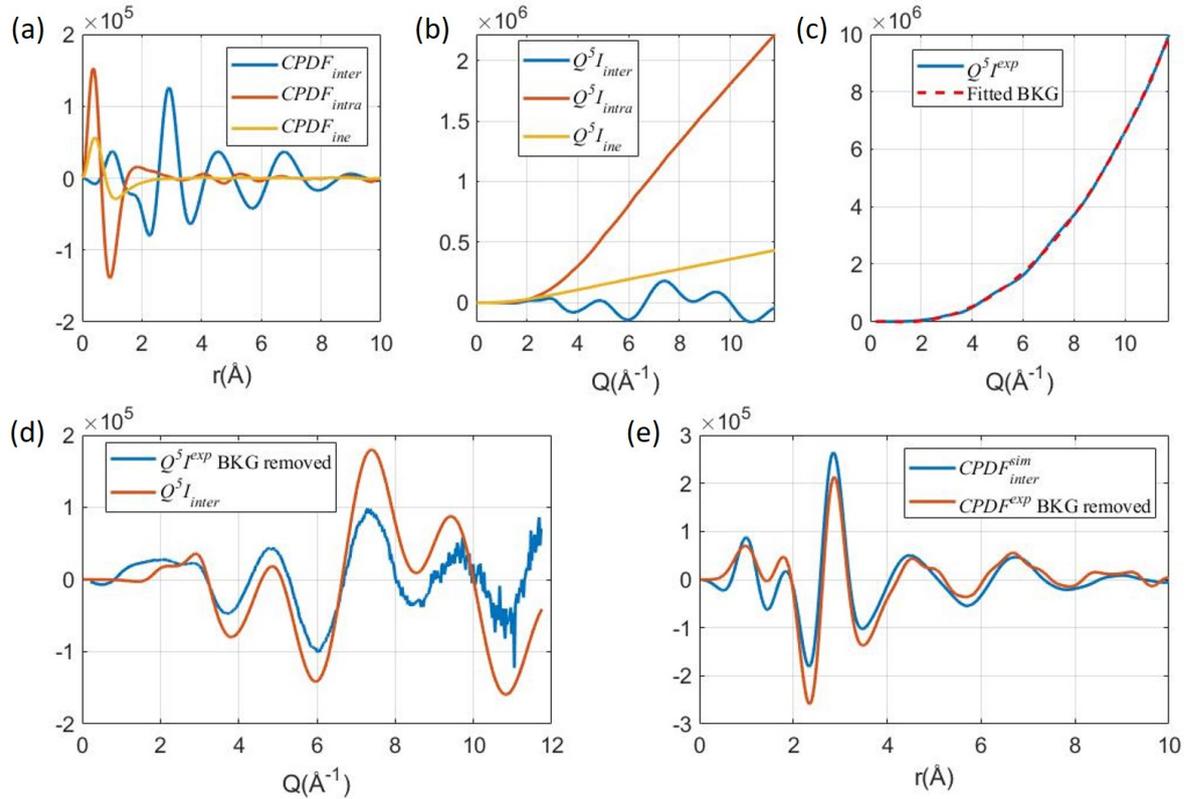

FIG. 3. Detailed analysis on liquid water. (a). The interatomic elastic $CPDF_{inter}$, intraatomic elastic $CPDF_{intra}$ and inelastic $CPDF_{ine}$ term of *CPDF* for water simulated under the IAM with a damping factor $\alpha = 0.06$. (b). Scattering pattern $Q^5I_{inter}$, $Q^5I_{intra}$, and $Q^5I_{ine}$ corresponds to the three terms shown in part (a). (c). A 3rd order polynomial fitting for background removal in experimental $Q^5I$. (d). The background removed experimental data compare with the simulated $Q^5I_{inter}$. (e) The $CPDF_{inter}$ from simulation (blue) and experimental *CPDF* without background removal (red), calculated using a damping factor $\alpha = 0.03$. All plots use $Q_{max} =11.8$ Å$^{-1}$. Experimental data is acquired using 100 nm water sheet jet.

## B. Experimental data for four liquids

In order to test the generality and robustness of the data treatment method, we have performed LES experiments and *CPDF* analyses on four liquid solvents: $H_2O$, $CCl_4$, $CHCl_3$, and $CH_2Cl_2$, delivered using a gas-accelerated liquid jet. The electron transmission for the four datasets are 88%, 57%, 35%, and 68%, respectively, comprising a wide range of liquid sheet conditions. The thickness of the liquid sheet for $CCl_4$, $CHCl_3$, $CH_2Cl_2$ was not directly measured, but has been estimated to be between 100 and 200 nm based on scattering cross-section, number density, and measured electron transmission. Fig. 4 shows the radially averaged raw scattering pattern $I(Q)$, the *CPDF* without background removal, and a graphic representation of the four liquid solvents, along with theoretical $I(Q)$ an *CPDF(r)* calculated from classical MD simulations. The experimental *CPDF* is calculated using Eq. (19). The structure of the four liquid solvents was simulated *via* classical MD with 50 × 50 × 50 Å boxes using the GROMACS software suite. Water was modelled using the TIP4P-Ew force field, and $CCl_4$, $CHCl_3$, $CH_2Cl_2$ were modeled using the OPLS-AA force field. The classical MD trajectories were transformed into time-averaged PDFs using the VMD package, and the scattering pattern simulation was carried out using the method of Dohn *et al.*[29] under the IAM. For each sample, five temperatures (270 K, 285 K, 300 K, 315 K, 330 K, and 345 K) were simulated, and the best fits (270 K for $CCl_4$, $CHCl_3$, $CH_2Cl_2$ and 285 K for $H_2O$) are shown in Fig. 4. It is known that OPLS-AA force field is inadequate for treating the electrostatic interaction for halogen atoms [30]. The reader should note, therefore, that the classical MD simulations carried out for $CCl_4$, $CHCl_3$, and $CH_2Cl_2$ are not intended to provide any rigorous comparison with the experimental data but, rather, to serve as references for the general appearance of the corresponding *CPDF*.

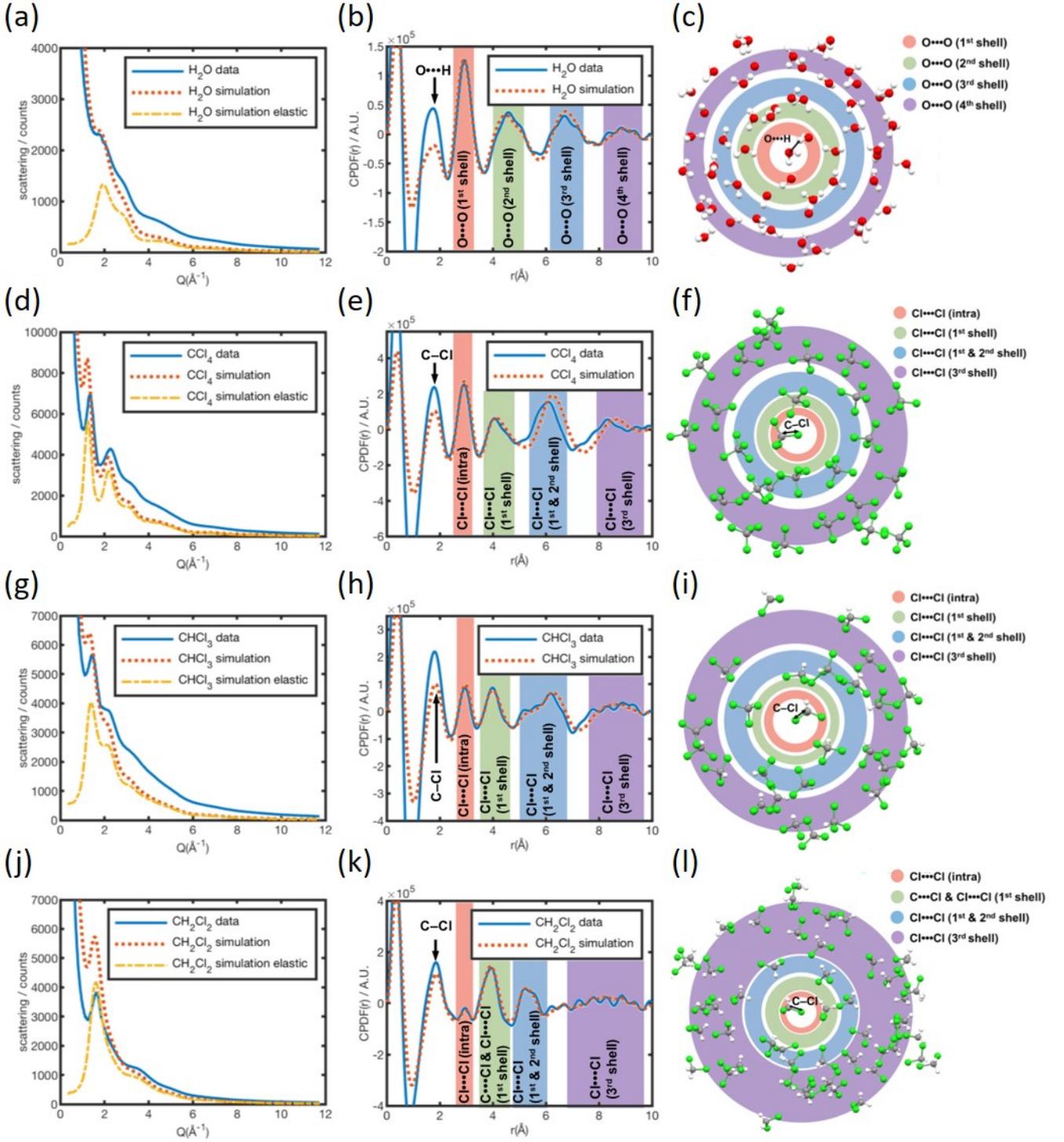

FIG. 4. Scattering pattern and *CPDF* for four different liquids. Radially-averaged electron scattering patterns (a, d, g, j), *CPDF* (b, e, h, k), and a graphic representation of atom pairs (c, f, i, l) of 4 liquids from experimental data and simulation. $Q_{max}$ =11.8 Å$^{-1}$ is used in both experimental and simulated patterns. A damping factor of $\alpha = 0.06$ is used for panel (b), (e), (h), $\alpha = 0.05$ is used for panel (k). All experimental data are taken with a gas-accelerated liquid jet.

## IV. DISCUSSION

### A. Understanding the *CPDF*

To better understand the information content of the *CPDF* and the interplay between nucleus-nucleus, electron-electron and nucleus-electron pair contribution, we simulated the *CPDF* of gas-phase CCl$_4$ molecules under the IAM. The elastic and inelastic atomic form factors are taken from ref. [31]. The charge pairs in the *CPDF* can be separated into three parts: elastic intraatomic contributions $CPDF_{intra}$, elastic interatomic contributions $CPDF_{inter}$, and inelastic contributions $CPDF_{ine}$, as shown in Fig. 5(a).

The elastic intraatomic contributions come from charge pairs within the same atom, and contains a large positive peak at $r = 0.4$ Å, a large negative peak at $r = 1$ Å and a small positive peak at $r = 2$ Å. The two positive peaks come from electron-electron pair and the negative peak represents electron-nucleus pair. The $CPDF_{intra}$ contribution dominates $r < 2$ Å, and becomes negligible at $r > 3$ Å, because intraatomic pairs are constrained to short distances. The physical origin of inelastic electron/X-ray scattering was shown by Bartell *et al.* to be electron correlation, *i.e.* electrons avoid each other spatially through Pauli exclusion and Coulomb repulsion [18]. The contribution $CPDF_{ine}$ therefore encodes a "correction" to the electron spatial distribution due to electron correlation. $P_{ine}$ becomes negligible at $r > 2$ Å, since electron correlation is also a short-ranged interaction. Therefore, longer-range ($r > \sim 2$ Å) interactions are dominated by the $P_{inter}$ term, in which structural information about the molecule is encoded. The current separation point ($r > \sim 2$ Å) is related to the maximum $Q$ range of this experiment. A detailed analysis of the influence of the $Q$ range on the *CPDF* analysis is given in the appendix (section VII C).

The $CPDF_{inter}$ contribution is plotted in Fig. 5(b) along with the nucleus-nucleus, electron-electron and nucleus-electron contributions. Despite the diverse nature of the various contributions, peaks in $CPDF_{inter}$ are found at $r = 1.72$ and $r = 2.92$ Å, demonstrating that $CPDF_{inter}$ encodes the C-Cl and Cl⋯Cl internuclear separation (1.77 and 2.89 Å, respectively) within 0.05 Å accuracy [32]. This net effect arises from the negative contributions from nucleus-electron pairs and the broad and weak nature of electron-electron pair contributions. Note that the electron-electron pair contribution (the yellow curve in Fig. 5(b)) is equivalent to the information content obtained from X-ray scattering but without applying AFF in the sine transformation.

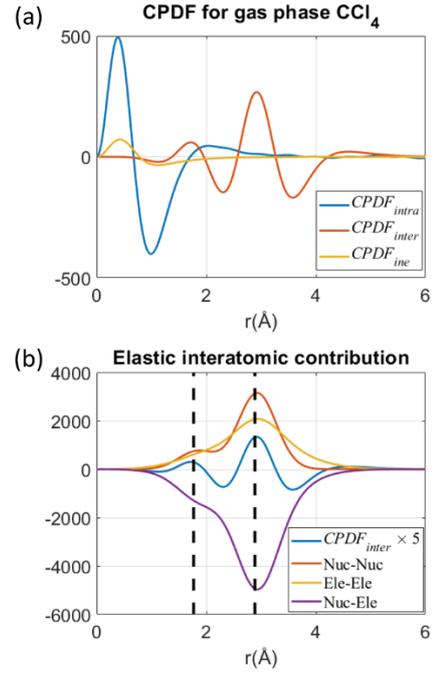

FIG. 5. Simulated *CPDF* for gas phase $CCl_4$ molecule. (a) The elastic intraatomic ($CPDF_{intra}$), elastic interatomic ($CPDF_{inter}$), and inelastic ($CPDF_{ine}$) contributions to *CPDF*. (b) $CPDF_{inter}$ with nucleus-nucleus, nucleus-electron, and electron-electron contributions. Scattering patterns are simulated under IAM, and *CPDF* are calculated using Eq. (19) in the main text with damping factor $\alpha = 0.06$ and $Q_{max} = 11.8$ Å$^{-1}$. The black dashed line shows the internuclear separation for C-Cl (1.77 Å) and Cl⋯Cl (2.89 Å), taken from ref. [32].

### B. Comparison to PDF

In our previous work, we showed the *PDF* analysis of LES from liquid water. Fig. 6 presents a direct comparison of *CPDF* and *PDF* analysis. While the *PDF* analysis published in our previous work [16] requires an empirical power fit for background removal and theoretical AFFs to calculate $sM$ (Fig. 6 (d-f)), the *CPDF* analysis only requires a $Q^5$ scaling (Fig. 6(a-c)). A comparison between Fig. 6(c) and (f) shows that both methods returned peaks corresponding to the 1$^{st}$ water shell (~2.9 Å), 2$^{nd}$ water shell (~4.5 Å), 3$^{rd}$ water shell (~6.8 Å). In addition, the *CPDF* also revealed a weak 4$^{th}$ water shell at ~8.9 Å, which can be attributed to the contrast enhancement provided by two negative shoulders beside each positive peak in *CPDF*.

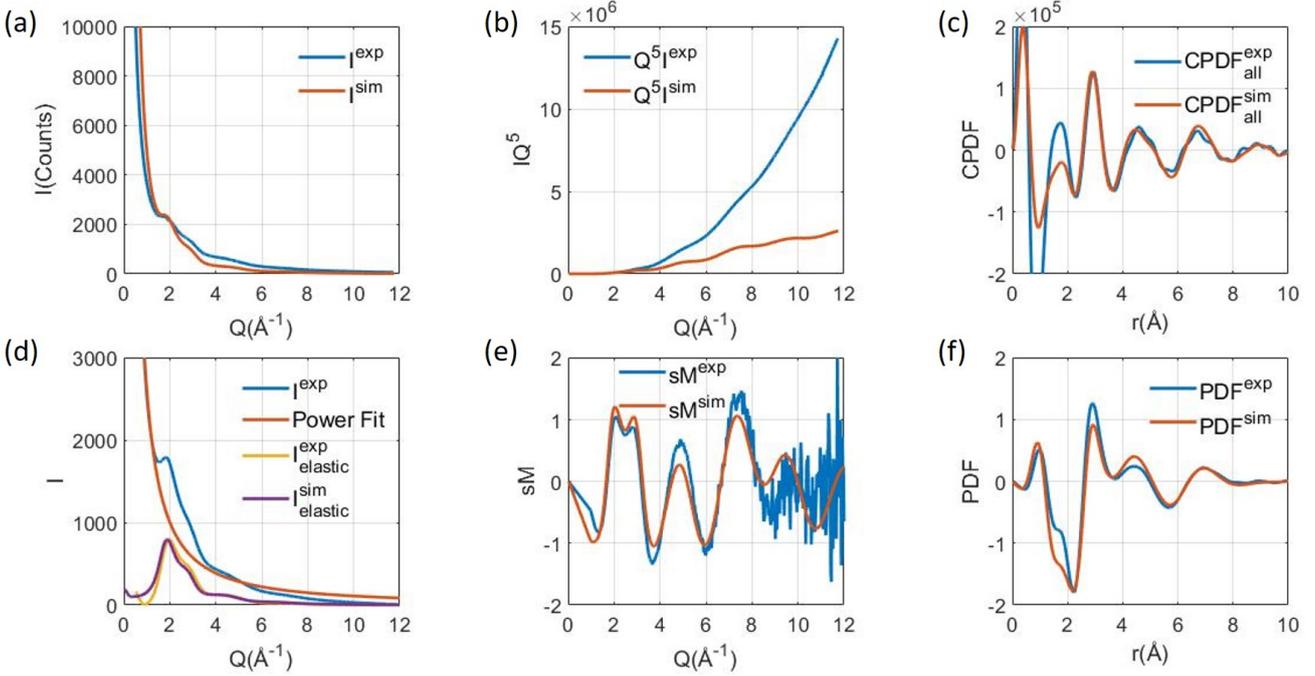

FIG. 6. Comparison of *CPDF* (a-c) and *PDF* (d-f) analysis on water. (a) Raw scattering intensity *I*, (b) $Q^5I$, (c) *CPDF*, (d) experimental raw scattering intensity, empirical power fit, elastic components obtained by removing the power fit, and simulated elastic component, (e) *sM*, (f) *PDF*. Exp, experimental; sim, simulated. Parts (d-f) are adopted from Nunes et al. [16].

## V. CONCLUSION

In summary, we propose *CPDF* as an alternative real-space analysis tool to the conventional *PDF* method for structure retrieval in liquid electron scattering experiments. *CPDF* is a mathematically rigorous transformation that does not invoke any theoretical model or empirical fitting. The generality and robustness of the *CPDF* method is demonstrated through the retrieval of major internuclear pairs between ~2 and ~10 Å in experimental LES data for $H_2O$, $CCl_4$, $CHCl_3$, and $CH_2Cl_2$. Although easily calculated, the interpretation of *CPDF* requires careful consideration as it overlays contribution form nucleus-nucleus pairs as sharp peaks, nucleus-electron pairs as negative shoulders on either side of each nucleus-nucleus pairs, and electron-electron pair contribution as broad and smooth features. *CPDFs* enable access to structural information from liquid-phase scattering directly through raw scattering patterns without the need of theoretical input or empirical fitting, thus making *CPDF* analysis a valuable tool in future LES studies.

## VI. ACKNOWLEDGEMENT

We gratefully acknowledge Dr. Xiaolei Zhu for the discussion that motivated this experiment. We thank Gregory M. Steward from SLAC National Accelerator Laboratory for making Figure 1A. The experiment was performed at SLAC MeV-UED, which is supported in part by the DOE BES SUF Division Accelerator & Detector R&D program, the LCLS Facility, and SLAC under contract Nos. DE-AC02-05-CH11231 and DE-AC02-76SF00515. J. P. F. N., and M. C. are supported by the U.S. Department of Energy Office of Science, Basic Energy Sciences under Award No. DE-SC0014170. K. L. is supported by a Melvin and Joan Lane Stanford Graduate Fellowship and a Stanford Physics Department fellowship. A. C. and T. J. A. W. are supported by the U.S. Department of Energy, Office of Science, Basic Energy Sciences, Chemical Sciences, Geosciences, and Biosciences Division.

## VII. APPENDIX

### A. Details of experimental setup

Experimental data is recorded at the SLAC MeV-UED facility. The detailed design and characterization of the experimental setup is reported in an earlier publication [16]. In brief terms, the electron beam is generated and accelerated by an rf-type photoinjector to a kinetic energy of 3.7 MeV. The pulse contains roughly 60,000 electrons per pulse at birth, and roughly 15,000 electrons per pulse at sample location. The electron beam size is 88 (H) × 37 μm (V) FWHM, measured by a knife-edge scan. The electron pulse has a repetition rate of 360 Hz.

A description of the two types of ultrathin liquid jet used in this work is also given in Ref. [16]. The key part of the gas-accelerated jet is a microfluidic chip with two gas channels and a liquid channel. High-pressure helium is used to flatten the liquid sheet at interaction region. The operating condition for datasets acquired in this work are as follows: $H_2O$—72 psi helium pressure, 0.20 mL/min flow rate; $CCl_4$—75 psi helium pressure, 0.20 mL/min flow rate; $CHCl_3$—80 psi helium pressure, 0.25 mL/min flow rate; $CH_2Cl_2$—77 psi helium pressure, 0.18 mL/min flow rate. The converging jet, also based on microfluidic chip technology, uses two liquid channels angled towards each other to produce a liquid sheet. In this work, only water is studied with the converging jet. The sample flow rate is 2.4 mL/min when operating the converging jet; the thickness of the converging jet is measured to be ~650 nm.

**B. Simulating *CPDF* contributions under IAM**

*CPDF* contains contribution from intraatomic, interatomic and inelastic components. For interatomic part, it is interesting to further separate it into nucleus-nucleus, nucleus-electron and electron-electron pairs, as shown in Fig. 3 and 5. This section introduces how each component is simulated.

In general, we first simulate the scattering pattern of each term, the *CPDF* is then calculated using Eq. (19) in the main text. Under IAM, the form factor for elastic X-ray scattering $F^X(Q)$ and inelastic X-ray scattering $S^X(Q)$ is tabulated in ref.[31]. The form factor for elastic and inelastic electron scattering, $F^E(Q)$ and $S^E(Q)$, can be written as [18, 33]

$$F^E(Q) = \frac{Z - F^X(Q)}{Q^2} \quad (20)$$

$$S^E(Q) = \frac{S^X(Q)}{Q^4} \quad (21)$$

The superscript *E* and *X* represents electron and X-ray scattering, respectively. *Z* is the nuclear charge of the atom. In Eq. (20), *Z* represents the elastic scattering from the nucleus, $F^X(Q)$ represents the elastic scattering from the electrons, and the denominator $Q^2$ comes from the $r^{-1}$ Coulomb potential [33].

The inelastic scattering pattern is simply the sum of inelastic scattering cross section of individual atoms

$$I^E_{ine}(Q) = \sum_{m=1}^{N} S^E_m(Q) \quad (22)$$

where the subscript *m* is summed over all atoms in the target system, and *N* is the total number of atoms. The elastic scattering pattern can be calculated using the following formula [33]

$$I^E_{ela}(Q) = I^E_{intra}(Q) + I^E_{inter}(Q) = \sum_{m=1}^{N} |F^E_m(Q)|^2 + \sum_{m=1}^{N}\sum_{n=1, n \neq m}^{N} |F^E_m(Q)||F^E_n(Q)|\frac{\sin(Qr_{mn})}{Qr_{mn}} \quad (23)$$

where $r_{mn}$ is the distance between the $m^{th}$ and the $n^{th}$ atom. Here the first term is intraatomic component and the second term is interatomic component. Re-writing Eq. (20) as:

$$F^E(Q) = F^{nuc}(Q) + F^{ele}(Q) \quad (24)$$

where

$$F^{nuc}(Q) = \frac{Z}{Q^2}, F^{ele}(Q) = -\frac{F^X(Q)}{Q^2} \quad (25)$$

The intermolecular component can then be written as:

$$I_{inter}^{E}(Q) = \sum_{m=1}^{N}\sum_{n=1,n\neq m}^{N}[F_m^{nuc}(Q)+F_m^{ele}(Q)][F_n^{nuc}(Q)+F_n^{ele}(Q)]\frac{\sin(Qr_{mn})}{Qr_{mn}}$$
$$= \sum_{m=1}^{N}\sum_{n=1,n\neq m}^{N}[F_m^{nuc}(Q)F_n^{nuc}(Q)+F_m^{ele}(Q)F_n^{ele}(Q)+F_m^{nuc}(Q)F_n^{ele}(Q)+F_n^{nuc}(Q)F_m^{ele}(Q)]\frac{\sin(Qr_{mn})}{Qr_{mn}} \quad (26)$$

where the first term $F_m^{nuc}(Q)F_n^{nuc}(Q)$ represents nuclear-nuclear pairs, the second term $F_m^{ele}(Q)F_n^{ele}(Q)$ represents electron-electron pairs, and the other two terms $F_m^{nuc}(Q)F_n^{ele}(Q)+F_n^{nuc}(Q)F_m^{ele}(Q)$ represent nuclear-electron pairs. Substituting Eq. (25) into Eq. (26) one can calculate each component separately.

For liquid phase samples, it is usually more efficient to directly simulate scattering patterns from $g_{mn}(r)$, the radial distribution function of atom pair $mn$. We use the method proposed by Dohn et al. [29]

$$I_{ela}^{E} = \sum_{m=1}^{N}F_m^{E}(Q)^2 + \sum_{m,n}F_m^{E}(Q)F_n^{E}(Q)\frac{N_m(N_n-\delta_{m,n})}{V}\int_0^R\frac{\sin(Qr)}{Qr}4\pi r^2[g_{mn}(r)-1]dr \quad (27)$$

where $\delta_{m,n}$ is the kronecker delta function, $V$ is the volume of the box in simulation, and $R$ is the largest distance in $g_{mn}(r)$.

### C. Influence of maximum $Q$ range on *CPDF*

In our LES experiment, we measure diffraction up to $Q_{max}$=11.8Å$^{-1}$, which is much smaller than the state-of-the-art for both neutron and X-ray scattering. Here we use the simulated liquid water scattering to show the impact of $Q$ range to the *CPDF* retrieval.

The interatomic, intraatomic and inelastic *CPDF* for $Q_{max}$=6, 9, 12, 15, 18, 21 Å$^{-1}$ are shown in Fig. 7. To avoid edge effects in the sine transform, the damping factor $\alpha$ is chosen so that $e^{-\alpha Q_{max}^2} = 2.3\times10^{-4}$, matching the case for this experiment ($\alpha$ = 0.06 and $Q_{max}$ =11.8 Å$^{-1}$). This choice of the damping factor treats the edge effect equally for various $Q_{max}$, but does introduce different level of peak broadening. For this reason, *CPDF*$_{intra}$ and *CPDF*$_{ine}$ extends to different $r$ range for different $Q_{max}$, and the hydrogen bond peak at ~1.9 Å is absent in *CPDF*$_{inter}$ for $Q_{max}$=6 Å$^{-1}$ and $Q_{max}$=9 Å$^{-1}$. For $Q_{max}$=6 Å$^{-1}$ the *CPDF*$_{inter}$ becomes the dominating term at roughly r=2.7 Å, while for $Q_{max}$=21 Å$^{-1}$ the *CPDF*$_{inter}$ becomes the dominating term at roughly r=1.3 Å. In the current experiment, this separation is at ~2 Å, right above the ~1.9 Å hydrogen bond peak, and a background removal is needed to reveal this peak. The simulations in Fig. 7 show that an experiment with a higher $Q$ range should be able to resolve the hydrogen bond peak directly without background removal. However, since the high $Q$ signal in the current experiment might be dominated by multi-scattering (Fig. 3B in the main text), this might require a thinner liquid sheet.

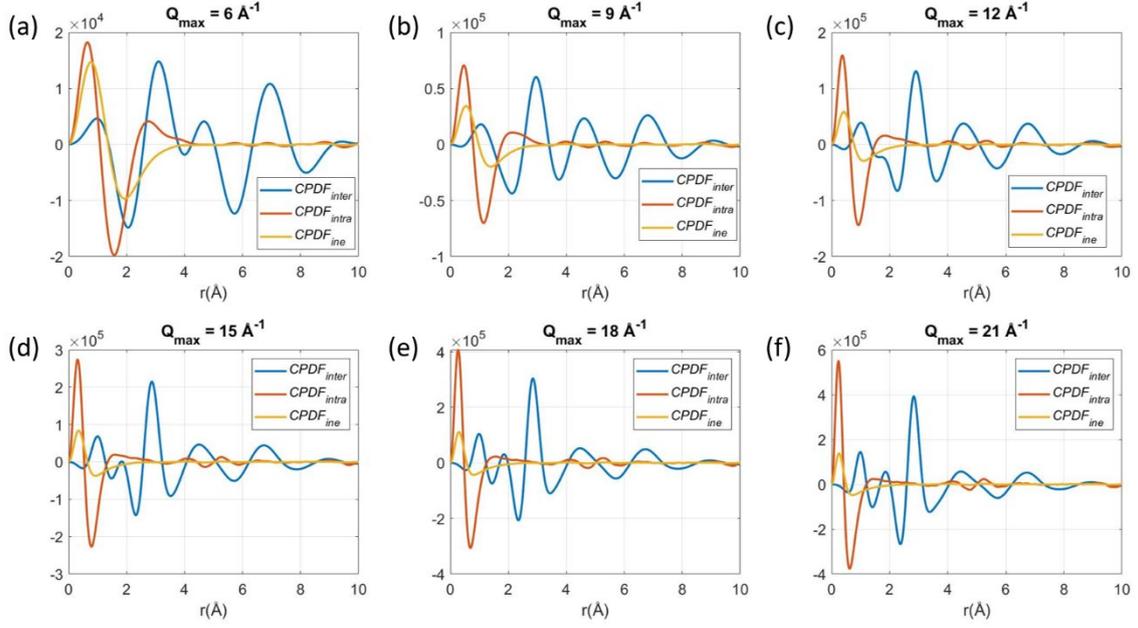

FIG. 7. Impact of $Q_{max}$ on *CPDF*. The *CPDF*$_{inter}$, *CPDF*$_{intra}$ and *CPDF*$_{ine}$ for 6 different maximum $Q$ ranges for $H_2O$, using simulated scattering patterns under IAM. The damping factor $\alpha$ is chosen so that $e^{-\alpha Q_{max}^2} = 2.3 \times 10^{-4}$ (see text).

### D. Simulated *CPDF*$_{inter}$ by pairs in all four liquids

Fig. 8 show the simulated *CPDF*$_{inter}$ by different atom pairs for all 4 liquids under GROMACS+IAM simulation. The dominating pairs shown in Table I is identified through this simulation.

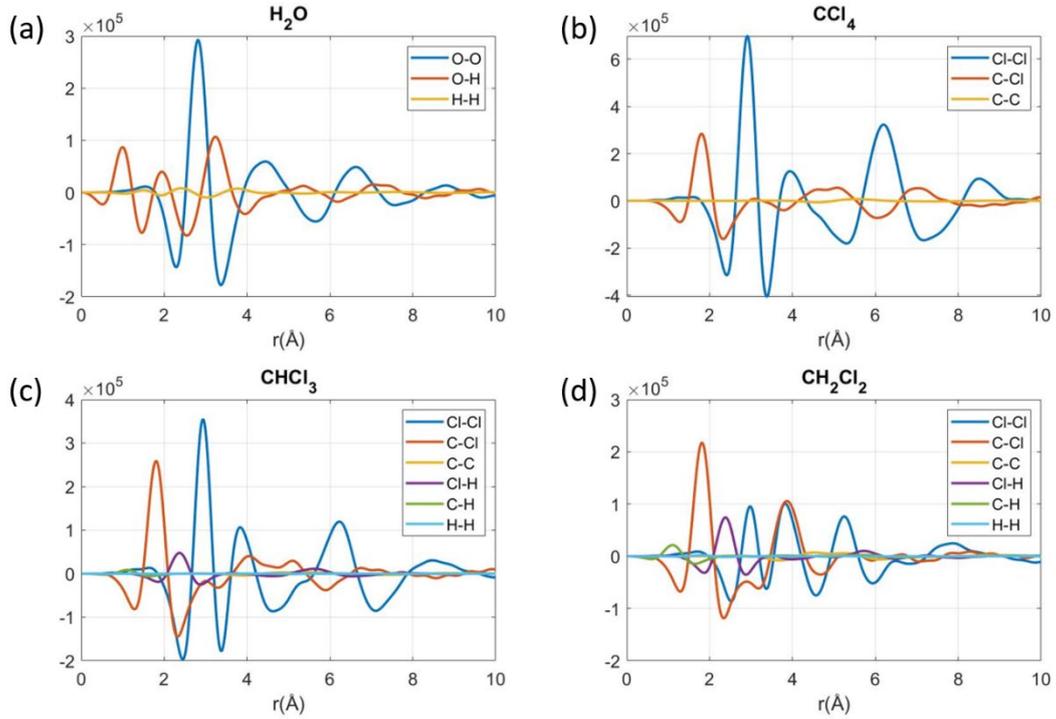

FIG. 8. Simulated $CPDF_{inter}$ by pairs for 4 liquids. Each panel gives the $CPDF_{inter}$ by each types of atom pairs, calculated using Eq. (19) in the main text with damping factor $\alpha = 0.03$ and $Q_{max} = 11.8$ Å$^{-1}$.